\newcommand\apjcls{1}
\newcommand\aastexcls{2}
\newcommand\othercls{3}

\newcommand\papercls{\aastexcls}
\documentclass[tighten,times,twocolumn]{aastex62}  

\if\papercls \apjcls
\usepackage{apjfonts}
\else\if\papercls \othercls
\usepackage{epsfig}
\usepackage{margin}
\usepackage{times}
\fi\fi
\usepackage{ifthen}
\usepackage{natbib}
\usepackage{bm}
\usepackage{amssymb, amsmath}
\usepackage{appendix}
\usepackage{etoolbox}
\usepackage[T1]{fontenc}
\usepackage{paralist}
\usepackage{newtxtext,newtxmath}
\if\papercls \apjcls
\newcommand\aas{\ref@jnl{AAS Meeting Abstracts}}
\newcommand\dps{\ref@jnl{AAS/DPS Meeting Abstracts}}
\newcommand\maps{\ref@jnl{MAPS}}
\else\if\papercls \othercls
\usepackage{astjnlabbrev-jh}
\fi\fi

\if\papercls \aastexcls
\hypersetup{citecolor=blue, 
            linkcolor=blue, 
            menucolor=blue, 
            urlcolor=blue}  
\else
\usepackage[
bookmarks=true,           
bookmarksnumbered=true,   
colorlinks=true,          
citecolor=blue,           
linkcolor=blue,           
menucolor=blue,           
urlcolor=blue,            
linkbordercolor={0 0 1},  
pdfborder={0 0 1},
frenchlinks=true]{hyperref}
\fi
\if\papercls \othercls

\else

\fi

\providecommand{\adsurl}[1]{\href{#1}{ADS}}

\makeatletter
\patchcmd{\NAT@citex}
  {\@citea\NAT@hyper@{%
     \NAT@nmfmt{\NAT@nm}%
     \hyper@natlinkbreak{\NAT@aysep\NAT@spacechar}{\@citeb\@extra@b@citeb}%
     \NAT@date}}
  {\@citea\NAT@nmfmt{\NAT@nm}%
   \NAT@aysep\NAT@spacechar\NAT@hyper@{\NAT@date}}{}{}

\patchcmd{\NAT@citex}
  {\@citea\NAT@hyper@{%
     \NAT@nmfmt{\NAT@nm}%
     \hyper@natlinkbreak{\NAT@spacechar\NAT@@open\if*#1*\else#1\NAT@spacechar\fi}%
       {\@citeb\@extra@b@citeb}%
     \NAT@date}}
  {\@citea\NAT@nmfmt{\NAT@nm}%
   \NAT@spacechar\NAT@@open\if*#1*\else#1\NAT@spacechar\fi\NAT@hyper@{\NAT@date}}
  {}{}
\makeatother

\makeatletter
\DeclareRobustCommand{\lowcase}[1]{\@lowcase#1\@nil}
\def\@lowcase#1\@nil{\if\relax#1\relax\else\MakeLowercase{#1}\fi}
\makeatother

\DeclareSymbolFont{UPM}{U}{eur}{m}{n}
\DeclareMathSymbol{\umu}{0}{UPM}{"16}
\let\oldumu=\umu
\renewcommand\umu{\ifmmode\oldumu\else\math{\oldumu}\fi}

\if\papercls \othercls

\else

\fi

\let\oldsim=\sim
\renewcommand\sim{\ifmmode\oldsim\else\math{\oldsim}\fi}
\let\oldpm=\pm
\renewcommand\pm{\ifmmode\oldpm\else\math{\oldpm}\fi}
\newcommand\by{\ifmmode\times\else\math{\times}\fi}

\newbox{\wdbox}
\renewcommand\c{\setbox\wdbox=\hbox{,}\hspace{\wd\wdbox}}
\renewcommand\i{\setbox\wdbox=\hbox{i}\hspace{\wd\wdbox}}

\newcount\timect
\newcount\hourct
\newcount\minct
\newcommand\now{\timect=\time \divide\timect by 60
         \hourct=\timect Cltiply\hourct by 60
         \minct=\time \advance\minct by -\hourct
         \number\timect:\ifnum \minct < 10 0\fi\number\minct}

\catcode`@=11

\newcommand\comment[1]{}

\newcommand\commenton{\catcode`\%=14}

\renewcommand\math[1]{$#1$}
\newcommand\mathshifton{\catcode`\$=3}

\let\atab=&
\newcommand\atabon{\catcode`\&=4}

\let\oldmsp=\sp
\let\oldmsb=\sb
\def\sp#1{\ifmmode
           \oldmsp{#1}%
         \else\strut\raise.85ex\hbox{\scriptsize #1}\fi}
\def\sb#1{\ifmmode
           \oldmsb{#1}%
         \else\strut\raise-.54ex\hbox{\scriptsize #1}\fi}
\newbox\@sp
\newbox\@sb
\def\sbp#1#2{\ifmmode%
           \oldmsb{#1}\oldmsp{#2}%
         \else
           \setbox\@sb=\hbox{\sb{#1}}%
           \setbox\@sp=\hbox{\sp{#2}}%
           \rlap{\copy\@sb}\copy\@sp
           \ifdim \wd\@sb >\wd\@sp
             \hskip -\wd\@sp \hskip \wd\@sb
           \fi
        \fi}
\def\msp#1{\ifmmode
           \oldmsp{#1}
         \else \math{\oldmsp{#1}}\fi}
\def\msb#1{\ifmmode
           \oldmsb{#1}
         \else \math{\oldmsb{#1}}\fi}

\def\supon{\catcode`\^=7}

\def\subon{\catcode`\_=8}

\def\supsubon{\supon \subon}

\newcommand\actcharon{\catcode`\~=13}

\newcommand\paramon{\catcode`\#=6}

\comment{And now to turn us totally on and off...}

\newcommand\reservedcharson{ \commenton  \mathshifton  \atabon  \supsubon 
                             \actcharon  \paramon}

\catcode`@=12
\reservedcharson

\if\papercls \apjcls

\else

\fi

\newcommand\chisq{\ifmmode{\chi\sp{2}}\else\math{\chi\sp{2}}\fi}
\newcommand\redchisq{\ifmmode{ \chi\sp{2}\sb{\rm red}}
                    \else\math{\chi\sp{2}\sb{\rm red}}\fi}
\newcommand\Teq{\ifmmode{T\sb{\rm eq}}\else$T$\sb{eq}\fi}
\newcommand\mjup{\ifmmode{M\sb{\rm Jup}}\else$M$\sb{Jup}\fi}
\newcommand\rjup{\ifmmode{R\sb{\rm Jup}}\else$R$\sb{Jup}\fi}
\newcommand\msun{\ifmmode{M\sb{\odot}}\else$M\sb{\odot}$\fi}
\newcommand\rsun{\ifmmode{R\sb{\odot}}\else$R\sb{\odot}$\fi}
\newcommand\mearth{\ifmmode{M\sb{\oplus}}\else$M\sb{\oplus}$\fi}
\newcommand\rearth{\ifmmode{R\sb{\oplus}}\else$R\sb{\oplus}$\fi}

\begin{document}

\title{An Energy Perspective of Core Erosion in Gas Giant Planets}

\author{J. R. Fuentes}
\affiliation{\rm Department of Applied Mathematics, University of Colorado Boulder, Boulder, CO 80309-0526, USA}
\author{Christopher R. Mankovich}
\affiliation{\rm Jet Propulsion Laboratory, California Institute of Technology, Pasadena, CA 91109, USA}

\author{Ankan Sur}
\affiliation{\rm Department of Astrophysical Sciences, Princeton University, 4 Ivy Lane, Princeton, NJ 08544, USA}

\begin{abstract}
Juno and Cassini have shown that Jupiter and Saturn likely contain extended gradients of heavy elements. Yet, how these gradients can survive over billions of years remains an open question. Classical convection theories predict rapid mixing and homogenization, which would erase such gradients on timescales far shorter than the planets' ages. To address this, we estimate the energy required to erode both dense and fuzzy cores, and compare it to what the planet can realistically supply. If the entire cooling budget is available to drive mixing, then even a compact core can, in principle, be destroyed. But if mixing is limited to the thermal energy near the core, which is another plausible scenario, the energy falls short. In that case, Jupiter can erode a fuzzy core by up to approximately $10~\mearth$, but a compact one remains intact. Saturn's core is more robust. Even in the fuzzy case, only about $1~\mearth$ is lost, and if the core is compact, erosion is negligible. The outcome depends sensitively on the assumed initial temperature and entropy profiles. Hotter and more superadiabatic interiors are more prone to mixing. We suggest that 3D simulations of convection driven from above, with realistic stratification and enough depth (i.e., many density scale heights) would be of great interest to further constrain the energy budget for core erosion.
\end{abstract} 


\section{Introduction}

Observations from the Juno \citep{Bolton2017} and Cassini \citep{Spilker2019} missions have revolutionized our understanding of Jupiter and Saturn, challenging the traditional view of their internal structures. Models that match Juno’s gravity measurements and Saturn’s ring seismology suggest that neither planet is uniformly mixed and that their cores appear to be ``fuzzy'', i.e., instead of dense, well-defined cores composed purely of heavy elements with sharp core-envelope boundaries, they contain a gradual, extended region enriched in heavy elements mixed with hydrogen and helium, which may span to half the planet’s radius or more \citep{Fuller2014, Bolton2017, Bolton2017b, Wahl2017, Mankovich_2021, Militzer2022, Miguel2022, Idini2022, Howard_2023}. It is important to emphasize that there is no unique interior model for the distribution of heavy elements. This is particularly true for Jupiter, where the gravity data constrain the density profile, not the composition, and the results depend on assumptions about layering, the equation of state, and other model choices. Many solutions are consistent with the data, including those with a small central core and a gradual enrichment in heavy elements over much of the interior \citep[see, e.g., Figure 2 in][]{Helled2022}. However, despite not having a unique and self-consistent view of the planet's interior, there is general consensus that heavy elements are spread over a large fraction of the planet's radius.

This conclusion has profound implications for the formation and evolution of gas giants. One possibility is that the planet formed with a more diffuse distribution of heavy elements than previously envisioned in the standard core-accretion scenario for planet formation \citep{Helled2017,Muller2020,stevenson_et_al_2022, Bodenheimer2025}. Alternatively, a fuzzy core could possibly arise from a giant impact between a large planetary embryo and the proto Jupiter/Saturn \citep{Liu2019}. However, the conditions required for such an impact (head-on collision, and $10\mearth$ impactor) are unlikely \citep{Helled2022}. Moreover, \cite{Meier2025} recently showed that even head-on collisions are unlikely to create Jupiter's fuzzy core, based on a combination of 3D N-body simulations, 3D SPH impact simulations, and thermal evolution models.
Another explanation is that a fuzzy core could naturally arise from the solubility of core material in metallic hydrogen. Density functional molecular dynamics simulations have shown that water ice, magnesium oxide, and iron dissolve under the extreme temperatures and pressures of gas giant interiors \citep{Wilson2012a,Wilson2012b,Wahl2013}, potentially leading to a gradual compositional gradient rather than a distinct core-envelope boundary.

Among the different channels that could lead to a fuzzy core, it remains unclear how such a compositional gradient can persist over evolutionary timescales, since any primordial gradient is expected to be eroded and mixed by the overlying convection zone shortly after formation \citep{Guillot2004, Vazan2015, Knierim2024}. This is the problem we investigate in this work. To the best of our knowledge, the first estimate of the core erosion rate was provided by \citet{Guillot2004}. They derived it from an order of magnitude calculation, assuming that all the thermal energy carried by convective motions is used to transport heavy elements upward, thereby increasing the planet’s gravitational potential energy. Under the assumption of an incompressible fluid, they estimated the core erosion rate as $\dot{M}_{\mathrm{core}} \approx - 0.3 L(R/GM)$, where $M$ and $R$ are the planet’s mass and radius, respectively, $G$ is the gravitational constant, and $L$ is the planet’s surface luminosity. Using a 1D evolution model for Jupiter and Saturn, \citet{Guillot2004} estimated that over 4.6 Gyr, convection can erode approximately $20~M_{\oplus}$ from Jupiter’s core but only about $2~M_{\oplus}$ from Saturn’s core. Although these estimates are plausible based on energetic arguments, they have yet to be validated through laboratory experiments or numerical simulations.

Later, \cite{Moll2017} investigated core erosion under the assumption that the core is semiconvective, consisting of multiple convective layers separated by double-diffusive interfaces, forming a convective staircase. In their model, they proposed a new expression for the erosion rate, given by $\dot{M}_{\mathrm{core}} \propto \alpha L_{\mathrm{core}}/c_P$, where $\alpha$ is the coefficient of thermal expansion, $c_P$ is the specific heat capacity at constant pressure, and $L_{\mathrm{core}}$ is the luminosity at the core-envelope interface. The proportionality constant is of order unity and depends on the vertical transport of energy and composition across the staircase. Unlike the previous estimates by \cite{Guillot2004}, \cite{Moll2017} concluded that due to the much smaller luminosity near the core of the planet, only a few $M_{\oplus}$ would be eroded from Jupiter's core during its entire evolution. However, their model assumes that a convective staircase can persist throughout the entire planet's evolution. This assumption is uncertain, as all available numerical simulations in the literature indicate that convective staircases fully mix on short timescales \citep[see, e.g.,][]{Wood2013,Garaud2018,Fuentes2022, Tulekeyev2024}. Incorporating layer mergers into their model will enhance the transport through the staircase and change significantly their conclusions.

Recently, \cite{Helled2022} argued that, although the first law of thermodynamics allows for the possibility that convection could mix the entire planet--since the total gravitational plus internal energy can decrease (become more negative) as the planet transitions from an initial state with a larger radius and centrally concentrated heavy element core to a state with a smaller radius but homogeneous interior--the significantly lower heat capacity of heavy elements compared to that of hydrogen (by a factor of $\sim$ 20) means that the deeper interior lacks sufficient energy to mix the core.

The energetics of core erosion in gas giants, as summarized above, appear to be insufficiently well-known or accepted within the fluid dynamics community. Given the rapid advancements in computational capabilities and the increasing ability to conduct simulations of convection under more realistic conditions, it is timely to revisit the problem of core erosion and discuss how fluid simulations can contribute to our understanding of gas giant interiors. In Section~\ref{sec:energy}, we first review the energy estimates of \cite{Helled2022} on the erosion of a compact core and then extend these calculations to more realistic models. In Section~\ref{sec:erosion} we present core erosion rates for Jupiter and Saturn in light of the energy considerations discussed previously. Finally, we conclude in Section~\ref{sec:discussion} with a summary and discussion.

\section{Energy considerations: Order of Magnitude Analysis}\label{sec:energy}

In this section, we first review the energetics of core erosion as presented by \cite{Helled2022}, who proposed two potential scenarios for the mixing of a \emph{compact} and \emph{incompressible} core: a ``global'' scenario, in which any plausible core could be fully mixed (at least for Jupiter), and a ``local'' scenario, where convective erosion is insufficient to mix the core. We then extend these calculations using polytropic interior models.

\subsection{Global Story}

In the global picture of core erosion, the planet cools from above and dense cold convective flows sink into the interior, causing turbulent entrainment of the core material into the envelope. This is possible provided that $\Delta E_{\mathrm{thermal}} > \Delta E_{\mathrm{potential}}$,  where $\Delta E_{\mathrm{thermal}}$ represents the decrease in the planet’s thermal energy over time, while $\Delta E_{\mathrm{potential}}$ corresponds to the gravitational energy required for the upward transport of heavy elements (rock and ice).  The thermal energy released during cooling can be estimated as $\Delta E_{\mathrm{thermal}} \sim M \Delta T c_P$, where $M$ is the planet’s mass, $\Delta T$ is the average temperature decrease during its evolution, and $c_P$ is the specific heat capacity at constant pressure. Since the temperature scale height deep within the planet is comparable to the planet’s radius, we have $c_P / (\alpha g) \sim R$, where $\alpha$ is the coefficient of thermal expansion and $g$ is the gravitational acceleration. Thus, $\Delta E_{\mathrm{thermal}} \sim M \Delta T \alpha g R$. 

Following \cite{Guillot2004}, the energy required to redistribute a small, compact core of mass $M_{\mathrm{core}}$ within a planet of total mass $M$ and radius $R$ can be calculated as the difference in gravitational potential energy between two configurations: (1) a two-layer sphere, and (2) a sphere of uniform density. In the first configuration, the core has a density $\rho_{\mathrm{core}}$, extends up to mass $M_{\mathrm{core}}$ and radius $R_{\mathrm{core}}$, and is surrounded by an envelope with density $\rho_{\mathrm{envelope}}$, extending up to mass $M$ and radius $R$.  In the limit of a small core, where  $M_{\mathrm{core}}/M \ll 1$, $R_{\mathrm{core}}/R \ll 1$, and $\rho_{\mathrm{envelope}}/\rho_{\mathrm{core}}\ll 1$

\begin{gather}
E_i \sim -\dfrac{3}{5}\dfrac{GM^2}{R}\left(1 + \dfrac{1}{2}\dfrac{M_{\mathrm{core}}}{M}\right)~.
\end{gather}
The final state has $E_f = -(3/5)GM^2/R$. Therefore the energy needed is $\Delta E_{\mathrm{potential}} \sim (3/10)(GM/R)M_{\mathrm{core}} \sim 0.3 M_{\mathrm{core}} g R$. Plugging in some realistic numbers for Jupiter, $R\sim 7\times 10^9~\mathrm{cm}$, $M\sim 2\times 10^{30}~\mathrm{g}$,~ $c_P \sim 2\times 10^{8}~\mathrm{erg~g^{-1}~K^{-1}}$, $g\sim 3000~\mathrm{cm~s^{-2}}$ yields

\begin{gather}
\Delta E_{\mathrm{thermal}} \sim 4\times 10^{42}~\mathrm{erg}~\left(\dfrac{\Delta T}{10^4~\mathrm{K}}\right), \\ 
\Delta E_{\mathrm{potential}} \sim  7\times 10^{41}~\mathrm{erg}~\left(\dfrac{M_{\mathrm{core}}}{20~M_{\oplus}}\right).
\end{gather}
i.e., $\Delta E_{\mathrm{thermal}} > \Delta E_{\mathrm{potential}}$. This is because the thermal energy lost during the early evolution of Jupiter (approximately the first $10^8~\mathrm{yr}$ after formation) is substantial, provided the planet starts out sufficiently hot (e.g., central temperatures of a few $10^4~\mathrm{K}$ or $10^5$~K). Yet convective mixing is inherently a local process, and it remains unclear whether convection can dredge up material from depths many density scale heights below the region where energy is lost (due to cooling to space at the planet’s surface).

\subsection{Local Story}

Consider now the local picture. For convection to operate at any radius, the buoyancy flux due to thermal buoyancy must be larger than that associated to compositional buoyancy. At small radii, the heat flux and the associated thermal buoyancy are significantly reduced, and the gravitational acceleration is also lower. Then, this local picture says that the energy available to mix a core upward is the excess thermal energy in the core itself

\begin{equation}
\Delta E_{\mathrm{thermal,loc}} \sim c_{P,\mathrm{core}} M_{\mathrm{core}} \Delta T_{\mathrm{core}}~,
\end{equation}
where $\Delta T_{\mathrm{core}}$ represents the decrease of the core temperature. For a mixture of rock and ice, the specific heat is about 20 times smaller than for hydrogen, $c_{P,\mathrm{core}}\sim 10^7~\mathrm{erg~g^{-1}~K^{-1}}$. Therefore,
\begin{equation}\label{eq:E_core}
\Delta E_{\mathrm{thermal,loc}} \sim 10^{40}~\mathrm{erg}~\left(\dfrac{M_{\mathrm{core}}}{20~M_{\oplus}}\right)\left(\dfrac{\Delta T_{\mathrm{core}}}{10^4~\mathrm{K}}\right) < \Delta E_{\mathrm{potential}},
\end{equation}
i.e., the thermal energy available is not enough to mix the core upwards. Note that even for an initially hotter interior with $\Delta T_{\mathrm{core}} \sim 10^5~\mathrm{K}$, the $\Delta E_{\mathrm{thermal,loc}}$ would still be smaller than $\Delta E_{\mathrm{potential}}$ by almost an order of magnitude. Clearly, the global and local pictures are in tension with one another.

\subsection{Energy Requirements for Polytropic Interior Models} \label{sec:new_models}

The arguments presented above are based on fundamental principles but rely on oversimplified assumptions, such as treating the entire planet as incompressible (constant density) and concentrating heavy elements within a small core. To more accurately assess the energy requirements for core erosion, we extend the previous calculations using interior models constructed with different distributions of heavy elements.

For simplicity, and to avoid uncertainties associated with specific equations of state for hydrogen–helium mixtures, we follow the approach of \cite{Stevenson2020} and \cite{Idini2022}, adopting a polytropic model with index $n=1$. We then use volume additivity to express the density $\rho$ of a hydrogen--helium fluid and heavy elements as a function of pressure $P$

\begin{gather}
\dfrac{1}{\rho} = \dfrac{1-Z}{\sqrt{P/K}} + \dfrac{Z}{\rho_Z}~, \label{eq:vol_aditivity}
\end{gather}
where $\rho_Z$ is the density of heavy elements, $Z$ is the mass fraction of heavy elements, and $K \approx 2.1 \times 10^{12} ~\mathrm{cgs}$ represents the bulk elastic properties of the mixture. Note that Equation~\eqref{eq:vol_aditivity} can be rewritten as

\begin{gather}
P = K\rho^2 f^2~,
\end{gather}
where the function $f$ is

\begin{gather}
f(r) = \dfrac{1-Z}{1-\frac{\rho}{\rho_Z}Z}~.
\end{gather}
Treating $\rho$ as the density of a pure hydrogen-helium fluid and $\rho_Z$ as the density of ``rocky'' heavy elements \citep{Hubbard1989}, we approximate $\rho/\rho_Z \approx 0.42 (j_0(kr))^{0.6}$, where $k = (2\pi G/K)^{1/2}$, and $j_0(kr) = \sin(kr)/kr$ is a Bessel function of the first kind of order zero. Provided the radial distribution of heavy elements, $Z(r)$, the density profile of the mixture can be obtained from solving the equation of hydrostatic balance with boundary conditions $\rho = \rho_c$ and $\partial \rho/\partial r = 0$ at the center.

\begin{figure}
    \centering
    \includegraphics[width=\columnwidth
]{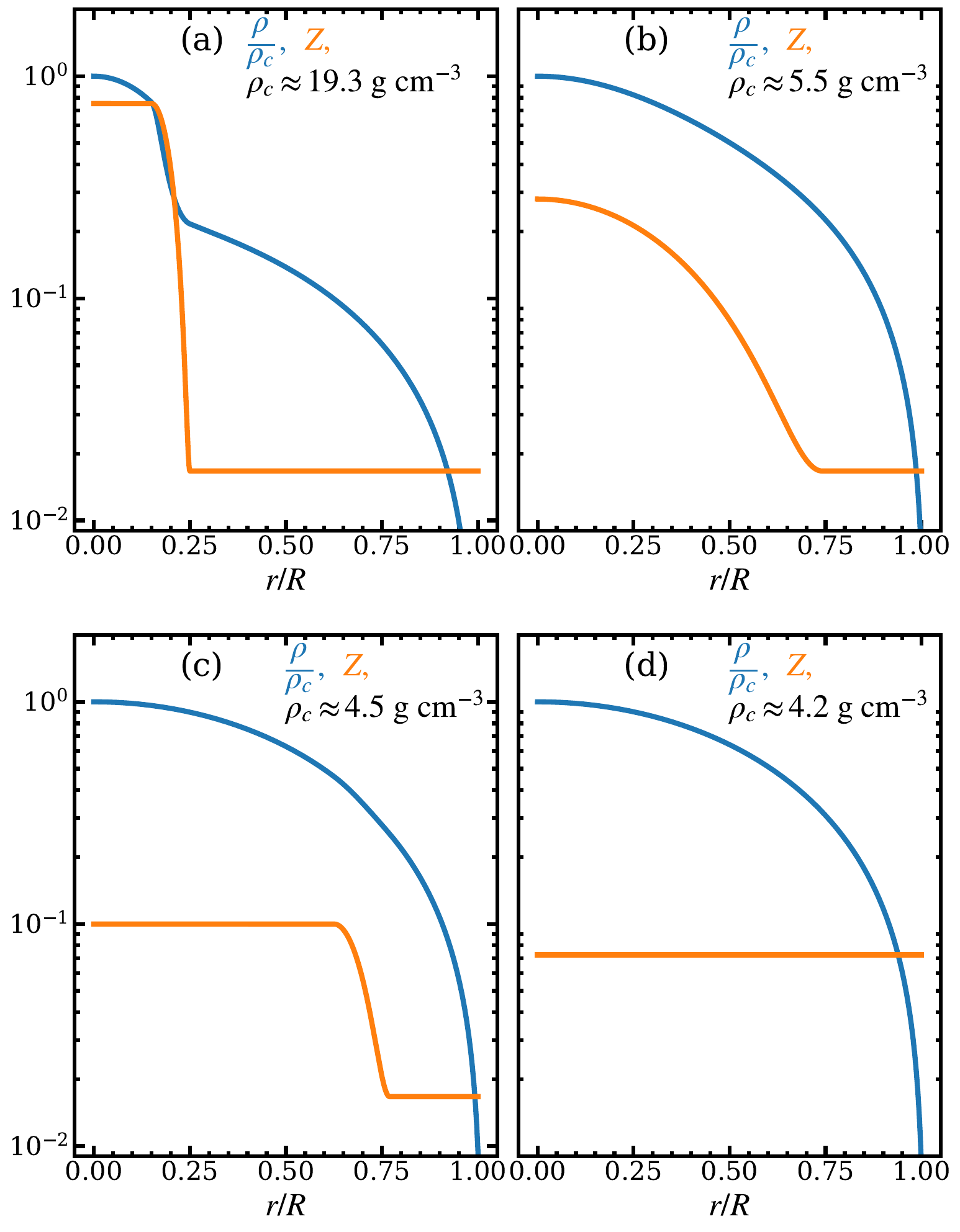}
    \caption{Interior models with different radial distributions of heavy elements. The blue and orange curves represent the density, normalized by the central density, and the mass fraction of heavy elements, respectively. Panel (a) shows a model with a narrow fuzzy core, where a sharp compositional gradient is confined to the central region. Panel (b) presents a fuzzy core with a smooth compositional gradient, similar to the structure proposed by \cite{Debras2019}. Panel (c) depicts a broad, dilute core with a sharp compositional gradient, akin to the model in \cite{Militzer2022}. Panel (d) corresponds to a fully mixed planet. In all models, the total mass of heavy elements within the gradient is fixed at $20 M_{\oplus}$.}
    \label{fig:interior_models}
\end{figure}

Figure~\ref{fig:interior_models} presents various interior models for a ``canonical'' gas giant, each with the same total heavy-element mass, 
$M_Z \sim 20~M_{\oplus}$, but differing in how these heavy elements are radially distributed within the planet. Assuming that models (a), (b), and (c) represent initial states, while model (d) is the final (homogeneous) state, we calculated the gravitational potential energy required to fully mix the heavy element distribution $Z(r)$. The energy changes for transitions from model (a) to (d), (b) to (d), and (c) to (d) are approximately $\Delta E_{\mathrm{potential}} \sim 10^{42}$ erg, $4.5 \times 10^{41}$ erg, and $2 \times 10^{41}$ erg, respectively. As expected, a less compact core requires less energy to redistribute heavy elements. Our previous estimates of the available thermal energy in the core (Equation~\ref{eq:E_core}) suggest that for $\Delta T_{\mathrm{core}} = 10^4~\mathrm{K}$, the thermal energy is approximately $10^{40}$ erg, making the complete erosion of the core challenging. However, for an initially hotter interior with $\Delta T_{\mathrm{core}} = 10^5~\mathrm{K}$, the available thermal energy increases to $10^{41}$ erg, suggesting that less compact cores could undergo substantial mixing.

\section{Core Erosion from evolution models} \label{sec:erosion}

In this section, we investigate core erosion using realistic evolution models of gas giants. In principle, the amount of mass mixed into the envelope can be estimated directly from the simulations. However, since evolution models rely on mixing length theory and lack a detailed transport model for convective mixing and entrainment, we adopt the approach taken by \cite{Moll2017}, who combined transport properties from 3D hydrodynamical simulations and detailed thermodynamic properties from 1D cooling models.

\subsection{Buoyancy flux ratio}
Of particular interest in the context of compositional mixing and entrainment of core material into the envelope is the ratio of the buoyancy flux associated with chemical transport across the interface, $F_C$, to the equivalent buoyancy flux associated with heat transport $F_T$

\begin{equation}
\gamma^{-1} \equiv \dfrac{\beta g F_C}{\alpha g F_T}~,
\end{equation}
where $\beta$ is the coefficient of solute contraction. Since the local luminosity at the core-evelope interface is given by $L_{\mathrm{core}} = 4\pi r_{\mathrm{core}}^2 \rho_{\mathrm{core}} c_P F_T$, where $r_{\mathrm{core}}$ is the core radius, and $\rho_{\mathrm{core}}$ is the local density at $r = r_{\mathrm{core}}$, and the mass loss rate of the core is $dM_{\mathrm{core}}/dt = -4\pi r_{\mathrm{core}}^2 \rho_{\mathrm{core}} \beta F_C$, we can express the core erosion rate as
\begin{equation}
\dfrac{dM_{\mathrm{core}}}{dt} \sim -\gamma^{-1} \dfrac{\alpha L_\mathrm{core}}{c_P} ~. \label{eq:rate}
\end{equation}

Experiments and modeling efforts suggest that when the compositional jump across an interface is sufficiently large, as in an initially differentiated planet, $\gamma^{-1}$ remains constant, independent of the fluid’s thermal and compositional stratification \citep[see, e.g.,][]{Turner_1965,Fernando1989}. Based on experiments with salty and sugar-rich water, \cite{Linden1978} proposed that
\begin{equation} \gamma^{-1} \approx \left(\dfrac{\kappa_C}{\kappa_T}\right)^{1/2} \equiv \tau^{1/2}~, \label{eq:gamma_old} \end{equation}
where $\kappa_C$ and $\kappa_T$ are the compositional and thermal diffusivities. In Jupiter’s core, the diffusivity ratio $\tau$ is estimated to be $\sim 10^{-2}$, implying $\gamma^{-1} \approx 0.1$ \citep[see, e.g.,][]{Stevenson1982,Guillot2004}. However, convective mixing and entrainment differ significantly between geophysical and astrophysical flows. A key parameter governing fluid dynamics is the Prandtl number,

\begin{equation}
\mathrm{Pr} = \dfrac{\nu}{\kappa_T}~,
\end{equation}
where $\nu$ is the kinematic viscosity. While $\mathrm{Pr} \approx 7$ in water, it is $\sim 10^{-2}$ in gas giant interiors. \cite{Moll2017} conducted hydrodynamical simulations of core erosion in fluids with $\mathrm{Pr} \sim \tau \sim 0.03$–$0.3$, assuming a double-diffusive (semiconvective) interface between the core and envelope, and constrained the buoyancy flux ratio to

\begin{align} \tau\left(\dfrac{\mathrm{Pr}+1}{\mathrm{Pr}+\tau}\right) < \gamma^{-1} < 1~.
\end{align}
For Jovian conditions ($\mathrm{Pr} \sim \tau \sim 10^{-2}$), this yields a lower bound of $\gamma^{-1} \approx 0.5$, implying an erosion rate five times larger than predicted by the water-based model of Equation~\eqref{eq:gamma_old}. 

In Section~\ref{sec:cooling}, we present cooling models for Jupiter and Saturn, whose properties at the core-envelope interface are used to integrate Equation~\eqref{eq:rate} and obtain the amount of core mass mixed by the outer convection zone as a function of time.
Since the buoyancy flux ratio $\gamma^{-1}$ is highly uncertain, we treat it as a free parameter and perform the integration over the range $\gamma^{-1} = 0.5$ to 1.

\begin{figure*}
    \centering
    \includegraphics[width=1.\linewidth]{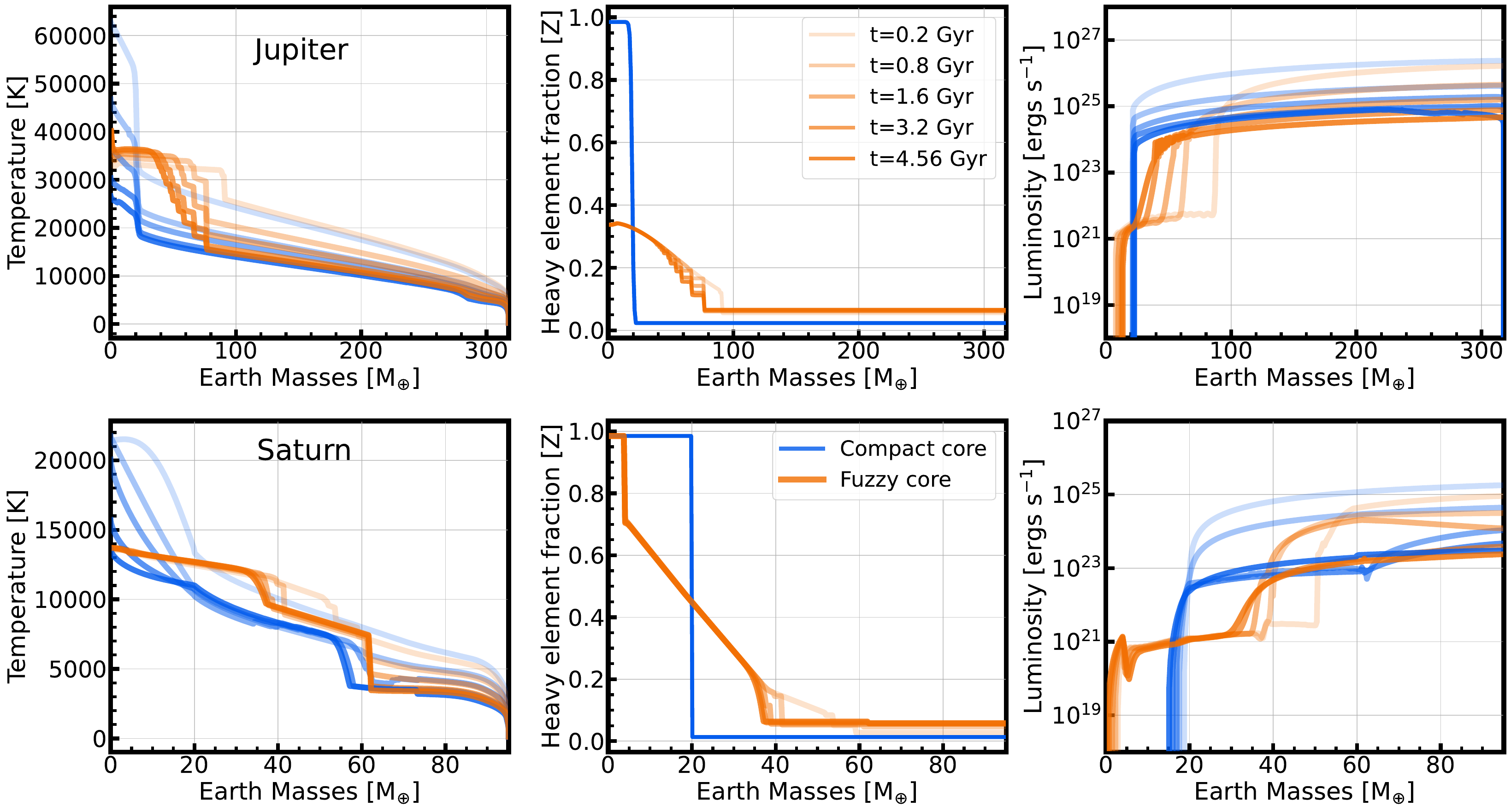}
    \caption{Evolutionary models for Jupiter (top row) and Saturn (bottom row) computed using the \texttt{APPLE} code \citep{Sur2024}. Blue curves denote compact‐core models and orange curves denote fuzzy‐core models (both with a 20 $M_\oplus$ core). Left panels: time evolution of the temperature profiles. Center panels: time evolution of the heavy‐element mass fraction $Z$. Right panels: luminosity versus time. All models are calibrated to match the observed effective temperatures of Jupiter and Saturn and Jupiter’s atmospheric helium abundance.}
    \label{fig:evolution}
\end{figure*}
\subsection{Cooling models} \label{sec:cooling}

We performed our cooling calculations for Jupiter and Saturn using the \texttt{APPLE} evolutionary code \citep{Sur2024}.  \texttt{APPLE} implements the \citet{Chabrier2021} hydrogen–helium equation of state (EOS) via the module of \citet{Arevalo2024}, applies the atmospheric boundary conditions of \citet{Chen2023}, and treats heavy elements with the AQUA EOS \citep{Haldemann2020}. Helium phase separation is modeled using the \citet[][L0911]{Lorenzen2011} miscibility curve. We use 400 mass zones for the Henyey iteration solver for all calculations.

In our compact-core models, we concentrate $M_{\rm core}\approx20 M_\oplus$ of heavy elements at the planetary center (so that $Z\approx1$ within the core), and distribute an additional $13 M_\oplus$ in Jupiter’s envelope and $1 M_\oplus$ in Saturn’s.
Convection dominates heat transport in the envelope according to the Ledoux criterion, while the compact core itself is assumed non-convective, losing heat solely by electron conduction.

For our fuzzy-core models, we build on the recent evolutionary calculations of \citet{Arevalo2025} and \citet{Sur2025a}, who first achieved simultaneous fits to both planets using fuzzy cores and the same microphysics. We slightly modify their initial $Z$ distributions to reassign mass between the core and envelope such that $M_{\rm core}\approx20$ \mearth, and apply the Ledoux criterion in regions of compositional stratification. To match Jupiter’s atmospheric helium abundance ($Y=0.234$; \citealt{vonZahn1998}), we introduce calibrated temperature shifts to the L0911 miscibility curve in both compact and fuzzy core scenarios. All our models reproduce the measured effective temperatures of 125.57 K for Jupiter \citep{Li2012}, and 96.67 K for Saturn \citep{Li2010}.

Figure~\ref{fig:evolution} presents results for Jupiter and Saturn based on two types of models: compact cores (blue curves) and fuzzy cores (orange curves). As expected, due to the much lower luminosity in the deep interiors of the planets, models with a compact core exhibit minimal mixing throughout the planet’s evolution. In contrast, models that begin with a fuzzy core undergo substantial mixing of core material into the outer convective envelope during the first $\sim 1~\mathrm{Gyr}$ of evolution. However, the innermost region of the primordial fuzzy core is preserved over time for two main reasons. First, the planet’s surface luminosity, which drives convection in the envelope, decreases by several orders of magnitude, significantly reducing the convective efficiency. Second, the base of the outer convection zone lies deep within the planet, where the buoyancy flux responsible for transporting heavy elements upward is weak due to both lower gravity and the reduced thermal heat capacity of the heavy elements. Still, we emphasize that the extent of mixing within the primordial fuzzy core is sensitive to initial conditions. Planets born with lower entropies are better at preserving their primordial interiors, while hotter starts favor greater erosion \citep{Knierim2024, Arevalo2025, Sur2025a}. For this reason, we deliberately adopt cooler initial states than those predicted by standard formation models \citep[e.g.,][]{Cumming2018,Muller2020,stevenson_et_al_2022}, which would otherwise drive more extensive mixing and erase much of the original core.

\subsection{Erosion rates}

\begin{figure}
    \centering
    \includegraphics[width=\columnwidth]{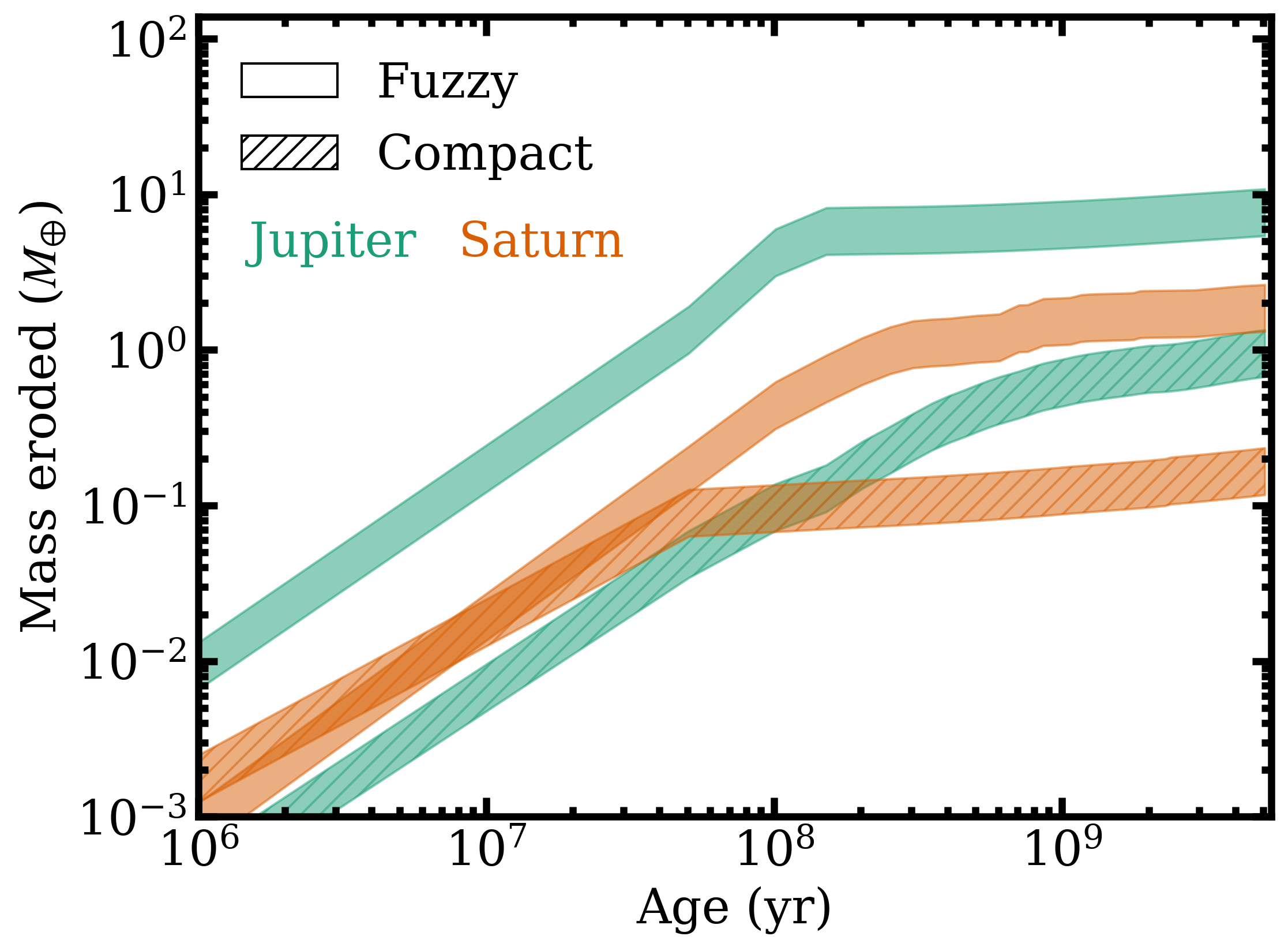}
    \caption{Core mass redistributed into the envelope as a function of time, using $\gamma^{-1}$ values between 0.5 and 1 (shaded areas). Results are presented for both Jupiter and Saturn, considering both compact and fuzzy core configurations.}
   \label{fig:erosion_rate}
\end{figure}

Since the erosion rate depends on properties evaluated at the core-envelope interface, we first identify the location of this interface by noting that the heavy-element mass fraction is approximately constant within the outer convective envelope. We define the boundary as the location where the heavy-element mass fraction first deviates by more than 5\% from its value within the convective zone, which corresponds to the last grid point of the convection zone, effectively marking its base. Although the 5\% threshold is somewhat arbitrary, it provides a reliable identification of the interface. Varying the threshold to 2.5\% or 10\% does not significantly affect our results.
Figure~\ref{fig:erosion_rate} shows estimates of the core mass eroded over time obtained by integrating Equation~\eqref{eq:rate} over a range of values of $\gamma^{-1}$. As expected from the ``local story'', in which the energy available for mixing comes from the thermal energy stored in the core, convective mixing remains limited for both Jupiter and Saturn, unless the core was initially fuzzy. For both planets, the core mass eroded by convection differs by roughly an order of magnitude between models with compact and fuzzy cores. In Jupiter, up to $\sim 10M_{\oplus}$ can be mixed into the envelope if the planet begins with a fuzzy core, compared to only $\sim 1M_{\oplus}$ for a compact core. In Saturn, the erosion rate is lower, also by about an order of magnitude, due to its reduced luminosity relative to Jupiter. In this case, up to $\sim 1M_{\oplus}$ can be mixed if the core is initially fuzzy, but only $\sim 0.1M_{\oplus}$ if it starts compact.

The results from the evolution models align well with the energy requirements discussed in Section~\ref{sec:new_models} for polytropic models, which demonstrate that fuzzy cores require less energy to undergo mixing. Also, the results suggest that the cores of both Jupiter and Saturn would likely be preserved today. This conclusion primarily arises from the use of the local luminosity at the core-envelope interface, $L_{\mathrm{core}}$, which is much smaller than the surface luminosity assumed in the global framework, where all thermal energy released during cooling is available for mixing the core.

\section{Summary and Discussion} \label{sec:discussion}

In this work, we investigated the energy requirements for convection to mix the core of a giant planet. We compared previous analytical estimates, which assumed an incompressible, dense core of small radius, with more realistic calculations based on two approaches: polytropic interior models derived from volume additivity and hydrostatic balance, and evolution models computed with \texttt{APPLE}. 

When estimating the energy requirements using polytropic interior models, we found that, within the global framework where energy released by cooling from above is used to mix the core, all cases should result in complete mixing of the dilute core, especially those that are less gravitationally bound (i.e., less compact). However, as noted by \citet{Helled2022}, a key open question in this context is whether cooling to space can effectively drive mixing at depths corresponding to roughly ten density scale heights below the surface. Addressing this challenge requires 3D simulations of penetrative convection in strongly stratified fluids, i.e., simulations that span many density scale heights, which lie beyond the validity of the Boussinesq approximation \citep{Spiegel_Veronis_1960}.

On the other hand, our evolutionary models of Jupiter and Saturn support the local scenario, in which the energy stored within the core itself acts as the reservoir for mixing, but this energy is insufficient to fully homogenize the core. In particular, mixing in Saturn is negligible for both fuzzy and compact cores, while in Jupiter, up to half of the core can be mixed if convection is efficient and the core starts out fuzzy. It would be 
instructive to quantify how different model assumptions, such as the shape of the internal structure, the adopted equations of state, and transport properties like thermal conductivities, affect the thermal energy available in the core. 

An important source of uncertainty is the planet's internal temperature. The core temperature of gas giants is not well constrained, but it could be significantly higher than $10^{4}~\mathrm{K}$. Because the planet's interior consists of a mixture of degenerate matter and ideal gas, estimating the central temperature from first principles is challenging. Nevertheless, for an ideal gas, the virial theorem suggests a characteristic central temperature of several $10^{5}~\mathrm{K}$, depending on the mean molecular weight. In this context, if the core temperature changes by a few $10^5~\mathrm{K}$, the total thermal energy available for mixing could be considerably larger than implied by Equation~\eqref{eq:E_core}. That said, evolution models with significantly hotter interiors could lead to a fully mixed planet, changing our conclusions \citep{Knierim2024, Arevalo2025}. This sensitivity to the internal temperature highlights the importance of the planet's thermal history. As shown by \citet{stevenson_et_al_2022} for Jupiter and more recently by \citet{Bodenheimer2025} for Saturn, the efficiency with which accretional energy is dissipated as heat during planet formation plays a crucial role in determining the planet's initial entropy and thermal structure. Improved formation models that better capture this process would therefore be of great interest.

There are several avenues that, if explored, could improve our understanding of core erosion in gas giants. For instance, the global picture may change considerably when additional physical processes that affect convective efficiency, such as rotation and magnetic fields, are taken into account. \citet{Hindman2023} employed an analytic entrainment model to show that rapid rotation, in combination with the declining surface cooling flux of a gas giant, can halt the growth of the outer convection zone. These analytic predictions have been recently confirmed by 3D numerical simulations (S. Zhang et al., in preparation). Such results are promising, as even rotation alone can substantially suppress the entrainment of heavy elements \citep{Fuentes2023} and delay the merger of semiconvective layers \citep{Fuentes2024,Fuentes2025}. Also, if the available energy in the core is indeed low, an important question is how semiconvection, usually invoked to explain why the core is not fully mixed \citep[e.g.,][]{Leconte_and_Chabrier_2012,Sur2025b}, could develop and be sustained over time. To assess whether the conditions necessary for double-diffusive convection are met and sustained within the core, it would be useful to have future work on first-principles estimates of the local density ratio, which is defined as
\begin{equation}
R_{\rho} \sim \frac{\beta \nabla Z}{\alpha (\nabla T - \nabla T_{\mathrm{ad}})},
\end{equation}
where $\nabla Z$ denotes the gradient of the heavy-element mass fraction, $\nabla T$ the thermal gradient, and $\nabla T_{\mathrm{ad}}$ the adiabatic gradient.
This is particularly relevant for Saturn, where seismology suggests the presence of a large, partially mixed fuzzy core that extends to about 60\% of the planet’s radius \citep[e.g.,][]{Mankovich_2021}. In light of our results, this implies that Saturn either formed with a fuzzy core or that substantial dilution (due to miscibility effects) of an initially compact core occurred early in its history. The fact that the core has not been mixed, highlights the importance of accurately calculating the apparently low mixing rate due to different physical processes acting in its interior.

Our results suggest that there may not be enough energy to fully mix the interiors of gas giants. Jupiter and Saturn are essential benchmarks for understanding the complex processes occurring in the interiors of giant exoplanets \citep[e.g.,][]{Bloot2023}. With the discovery of planets booming and the unprecedented atmospheric data provided by JWST, the assumption of a fully adiabatic and well-mixed interior in giant planets is becoming questionable. If these planets are not fully mixed, then the metallicities inferred from remote atmospheric observations do not accurately reflect their true interior compositions and planet’s bulk metallicity \citep{Howard2023,Muller2024}.

\begin{acknowledgements}
We thank the referee for providing a careful report that helped to improve the manuscript. We also thank Jim Fuller and Benjamin Idini for useful conversations on evolution models of gas giants. J.R.F. is supported by NASA Solar System Workings grant 80NSSC24K0927. C.M.’s research was supported by an appointment to the NASA Postdoctoral Program at the Jet Propulsion Laboratory, administered by Oak Ridge Associated Universities under contract with NASA. A.S. is supported by the Center for Matter at Atomic Pressures (CMAP), a National Science Foundation (NSF) Physics Frontier Center, under Award PHY-2020249. Any opinions, findings, conclusions, or recommendations expressed in this material are those of the author(s) and do not necessarily reflect those of the National Science Foundation. 
\end{acknowledgements}

\bibliography{references}{}
\bibliographystyle{aasjournal}

\end{document}